\documentclass[11pt]{article}
\usepackage{nicefrac}
\usepackage{amsfonts}
\usepackage{amssymb}
\usepackage{color}
\usepackage{amsmath,mathrsfs,dsfont}
\usepackage{graphicx}
\usepackage{appendix}
\usepackage{cite}
\newenvironment{proof}[1][Proof]{\noindent\textbf{#1.} }{\ \rule{0.5em}{0.5em}}

\newcommand{\Tr}{\mathrm{Tr}}
\makeatletter
\let\@fnsymbol\@arabic
\makeatother

\begin{document}

\title{On geodesics in space-times with a foliation structure: A spectral geometry approach.}
\author{A. Pinzul\thanks{apinzul@unb.br}\\
\\
\emph{Universidade de Bras\'{\i}lia}\\
\emph{Instituto de F\'{\i}sica}\\
\emph{70910-900, Bras\'{\i}lia, DF, Brasil}\\
\emph{and}\\
\emph{{International Center of Condensed Matter Physics} }\\
\emph{C.P. 04667, Brasilia, DF, Brazil} \\
}
\date{}
\maketitle

\begin{abstract}
Motivated by the Horava-Lifshitz type theories, we study the physical motion of matter coupled to a foliated geometry in non-diffeomorphism invariant way. We use the concept of a spectral action as a guiding principle in writing down the matter action. Based on the deformed Dirac operator compatible with the reduced symmetry - foliation preserving diffeomorphisms, this approach provides a natural generalization of the minimal coupling. Focusing on the IR version of the Dirac operator, we derive the physical motion of a test particle and discuss in what sense it still can be considered as a geodesic motion for some modified geometry. We show that the apparatus of non-commutative geometry could be very efficient in the study of matter coupled to the Horava-Lifshitz gravity.
\end{abstract}

\section{Introduction}

The absence of a definite theory of quantum gravity (QG) keeps the efforts in the direction of the construction of such a theory to be quite topical and very important. One of the recent proposals on a perturbatively renormalizable theory of QG is due to Horava \cite{Horava:2009uw}. The main idea of the Horava's approach (based on some older work by Lifshitz \cite{Lifshitz}, hence the name for such theories - Horava-Lifshitz (HL) gravity) is to give up the diffeomorphism invariance (Diff) on the fundamental level (so, it emerges as an effective symmetry in infrared (IR) regime only) in favor of the improved renormalizability of the theory. This is achieved by the inclusion of higher derivative (with respect to space coordinates) terms. As the immediate consequence, the space-time acquires the structure of a foliated manifold where space and time are separated and the resulting fundamental symmetry is given by foliation preserving diffeomorphisms (FPDiff) instead of full Diff, see \cite{Sotiriou:2010wn},\cite{Blas:2010hb},\cite{Mukohyama:2009zs} for some reviews on HL gravity. The presence of the higher order space derivatives as compared to the time derivative means that in the deep ultraviolet (UV) regime the theory becomes extremely non-relativistic with the anisotropic scaling for time, $t$, and space, $\vec{x}$ coordinates \cite{Horava:2009uw}
\begin{eqnarray}\label{scaling}
& &t \rightarrow a^z t  \nonumber\\
& &\vec{x} \rightarrow a\vec{x}  \nonumber \ .
\end{eqnarray}
$z$ is called the anisotropic scaling exponent. To make a theory of gravity at least formally renormalizable, $z$ should be not less then 3. The theory with $z=3$ is usually called the HL gravity.

So far we were talking only about pure gravity. The other side of the story is the coupling of matter to the anisotropic gravity \cite{Kimpton:2013zb}. The slightly oversimplified point of view on the problem is as follows: On one side, using Diff-breaking coupling in the matter sector will typically lead to the Lorentz breaking, which has very strong experimental bounds and, as the consequence, a lot of fine tuning should be used to meet those bounds; On the other side, using the minimal coupling as in General Relativity (GR) does not seem at all natural in a model based on FPDiff.\footnote{Though in this work we will not the explicit form of FPDiff, we will give it for the completeness and the convenience of a reader: $t'=t'(t)$, $\vec{x'}=\vec{x'}(t,\vec{x})$. So, the form of FPDiff explicitly demonstrates the existence of the preferred time.} A related point is that, in the usual construction, the gravitational part of the HL theory is not related to the matter part, except in sharing the same FPDiff symmetry. This gives the enormous freedom in writing both parts of the action each having a large number of independent parameters.

In this work, we suggest to use the spectral action principle \cite{Chamseddine:1996zu} as a way to relate this two independent parts. This potentially could address several important points: 1) reduce the number of free parameters; 2) relate parameters in the matter sector to those on the gravity side; 3) as the consequence, some fine tuning could be resolved in a natural way. In addition, this approach gives a more natural way of the matter-gravity coupling (which one can call minimal in some sense). The general idea is as follows (for the details see \cite{Chamseddine:1996zu},\cite{Chamseddine:2008zj}). One starts with some physically motivated Dirac operator, $\mathrm{D}$, defined on the Hilbert space of spinors $\psi$ and compatible with the symmetry of the system in hand (in our case FPDiff). Then one postulates that the full action, gravity plus matter, is given by
\begin{eqnarray}\label{fullalaction}
S_{spec} = \Tr f\left(\frac{\mathrm{D}^2}{\Lambda^2}\right)\ + \langle \psi|\mathrm{D}|\psi\rangle \ .
\end{eqnarray}
One can see that the same object, the Dirac operator, essentially defines both parts. This is the source of the possible advantages mentioned above. The approach based on (\ref{fullalaction}) has its roots in non-commutative geometry \cite{Connes:1994yd} and has been quite successfully used in the original approach to Standard Model \cite{Connes:2006qv}. The other areas where the ideas from non-commutative geometry have proven to be fruitful include, but not exhausted by, string theory \cite{Seiberg:1999vs}, black holes \cite{Dolan:2004xd}, entropic gravity \cite{Gregory:2012an}, etc.

Some time ago, we initiated the study of the HL-type models of gravity based on some natural deformation of Dirac operator and on active use of the methods of non-commutative geometry \cite{Pinzul:2010ct}. In the present paper, we apply these ideas to study the matter term in (\ref{fullalaction}). Because the analysis of the fully deformed Dirac operator is quite involved (this problem is currently under consideration), here we consider the particular limit of the full operator, in which it still remains of the first order in all derivatives (as in the case of the usual Dirac operator). Physically this can be interpreted as the IR limit of the full theory (as opposed to the UV regime, where all the higher derivatives are kept). Specifically, we study the following questions. What is the physical motion of a test particle in such a theory? What is the relation between this motion and the geodesics for the background geometry? How can we define either of them? All these questions are extremely important both conceptually and phenomenologically. At the concluding part, we are commenting on these points.

The structure of the paper is following. The section \ref{standard} is devoted to the development of our approach on the example of the Diff-invariant theory. We suggest an alternative way of deriving the physical motion of a test particle starting with the corresponding field theory. The approach is based on the Hamilton-Jacobi equation following from the quasiclassical analysis and treated as a relativistic Hamiltonian of a particle. Then we study the notion of a geodesic distance in the framework of non-commutative geometry showing that for a Diff-invariant theory it leads to the usual geodesics. In section \ref{anyzotropicgeodesic} we apply the introduced ideas to the case of a deformed theory based on FPDiff and study the question in what sense the physical motion still can be treated as a geodesic one. In the concluding section we briefly review the main results of the paper, as well as discuss the further steps, some of which are quite urgent to have the potential possibility of confronting the introduced ideas with experiment.

Also the paper has two appendices. While in the first one we set up the conventions and collect some definitions used in the main text, the second appendix on 3+1 decomposition of a Dirac operator has its independent value and should be useful in studies on FPDiff-invariant matter-gravity coupling.

\section{Geodesics: Standard case}\label{standard}

A motion of a test particle in GR is geodesic for the background pseudo-Riemannian geometry. It should be stressed that these two notions, physical motion of a test particle and geodesics, are \textit{a priori} unrelated. In the framework of GR, one can show that they coincide for the case of matter coupled to gravity in Diff-invariant way.\cite{Dixon:1970zz},\cite{Dixon:1970zza},\cite{Hawking:1973uf} This demonstration heavily relies on the existence of a covariantly conserved energy-momentum tensor (EMT), $T_{\mu\nu}=\frac{2}{\sqrt{-g}}\frac{\delta S_m}{\delta g^{\mu\nu}}$ ($S_m$ is the matter part of the full action). In HL-type theories based on FPDiff such a conserved tensor does not exist in general \cite{Kimpton:2010xi}. So, one would like to to have an alternative way to derive physical motion of a test particle starting with an action for a field coupled to gravity (which usually is more natural and fundamental then an action for a particle). While for the Diff-invariant coupling we should reproduce the usual equation for geodesic motion, the situation should change in the case of the FPDiff-invariant coupling.

The other point is the definition of a geodesic itself. For the case of the (pseudo)Riemannian geometry it is defined to be the extremal path between to points of the geometry. In a standard way, it can be found by extremizing the functional
\begin{eqnarray}\label{length}
L_{x_1 x_2}[s] = \int\limits_{x_1}^{x_2} ds \ ,
\end{eqnarray}
where, as usual, $ds^2 = g_{\mu\nu}dx^\mu dx^\nu$. This construction uses the information about the geometry in the form of its metric. What if we do not know what is the \textit{real} geometry of our theory? As we mentioned in introduction and will discuss more in the section \ref{anyzotropicgeodesic}, there exists a way of defining geometry in terms of a relevant Dirac operator. This seems to be a more physically motivated approach. So, we would like to have an alternative definition of a geodesic distance in terms of a Dirac operator. And indeed such a construction exists and produces the usual result in the case of the standard choice of a Dirac operator.

In this section, we discuss in details these two points, physical motion of a test particle and geodesics for a geometry defined by a Dirac operator, for the standard case of a Diff-invariant matter coupling and the usual Dirac operator. In the next section, we will generalize these constructions to the case of the FPDiff-invariant coupling and an appropriately deformed Dirac operator.

\subsection{Geodesics from the Hamilton-Jacobi eqation}\label{HJstandard}

Here we would like to derive the usual geodesic equation
\begin{eqnarray}\label{geodesic}
\frac{d^2 x^\mu}{d \tau^2}+\Gamma^\mu_{\nu\lambda}\frac{d x^\nu}{d \tau}\frac{d x^\lambda}{d \tau} = 0 \ ,\ \tau\ \mathrm{is\ a\ proper\ time}
\end{eqnarray}
starting with some diffeomorphism invariant field theory for matter. As we said earlier, while doing this, we do not want to appeal to the existence of the covariantly conserved energy-momentum tensor.

As we briefly discussed in the introduction, from the point of view of the spectral action, it is natural to work with the Dirac-type action (see the second term in (\ref{fullalaction})):
\begin{eqnarray}\label{action}
S_m = \int d^4 x \sqrt{-g}\,\bar{\psi}\mathrm{D}_m \psi \ ,
\end{eqnarray}
where $\mathrm{D}_m := \mathrm{D} - \frac{mc}{\hbar}$ and $\mathrm{D}=\gamma^\mu \nabla^\omega_\mu$ is the massless Dirac operator (\ref{Dirac}).\footnote{To allow for more smooth presentation of the main ideas and results, we collect in two appendices all the notations, conventions and some auxiliary calculations used in the main text.} $\psi (x)$ is a section of the spinor bundle over the space-time manifold $\mathcal{M}$, i.e. a spinor field. Also we explicitly keep track of $\hbar$ in view of the future quasiclassical analysis.

The equation of motion following from (\ref{action}) is the generally covariant form of the Dirac equation
\begin{eqnarray}\label{Diracequation}
\left(\gamma^\mu\nabla^\omega_\mu - \frac{mc}{\hbar}\right)\psi =0 \ .
\end{eqnarray}
In QFT, (\ref{Diracequation}) is the equation for the operator of a quantum relativistic spinor field, but being restricted to the 1-particle sector, this equation can be understood as a generalization of Schrodinger equation to the case of a relativistic spinor (exactly in the same way as Klein-Gordon equation for the case of a spin zero field). Then $\psi (x)$ is interpreted as the corresponding wave-function.

Now, we would like to look at the quasiclassical approximation of (\ref{Diracequation}). Toward this end, we write $\psi$ in a standard form
\begin{eqnarray}\label{quasiclassic1}
\psi = \chi \mathrm{e}^{\frac{i}{\hbar}S} \ ,
\end{eqnarray}
where $\chi$ is a 4-spinor, while $S$ is a scalar function. The reasoning for doing so is following: while $\bar{\chi}\chi$ is interpreted as the density of particles satisfying the continuity equation, $S$ is the quantum version of the Hamilton's principal function. The knowledge of this function (or, rather its zero order in $\hbar$ term) will allow us to read off the relativistic particle Hamiltonian following from this field theory, which will lead to the equation of motion for a particle.

Plugging(\ref{quasiclassic1}) into the equation of motion (\ref{Diracequation}), we get
\begin{eqnarray}\label{quasiclassicfermion}
\gamma^\mu\nabla^\omega_\mu \chi + \frac{i}{\hbar}\gamma^\mu \chi \partial_\mu S - \frac{mc}{\hbar}\chi = 0\ .
\end{eqnarray}
We are interested in the quasiclassical analysis of (\ref{quasiclassicfermion}). Writing the $\hbar$-expansion of $\chi$ and $S$
\begin{eqnarray}\label{quasiclassicexpansion}
\left\{
  \begin{array}{l}
    \chi = \chi_0 + \hbar\chi_1 + \cdots \ \\
    S = S_{cl} + \hbar S_1 + \cdots \\
  \end{array}
\right. \ ,
\end{eqnarray}
we have from the leading $\frac{1}{\hbar}$ term of (\ref{quasiclassicfermion})
\begin{eqnarray}\label{quasiclassicfermion1}
i\gamma^\mu \chi_0 \partial_\mu S_{cl} - mc\chi_0 = 0\ .
\end{eqnarray}
Here $S_{cl}$ is just a classical Hamilton's principal function, i.e. a classical action evaluated on a classical trajectory. It is more convenient to re-write (\ref{quasiclassicfermion1}) as to get rid of gamma-matrices and the spinor $\chi_0$. It is trivial to show that (\ref{quasiclassicfermion1}) is equivalent to
\begin{eqnarray}
(g^{\mu\nu}\partial_\mu S_{cl} \partial_\nu S + m^2 c^2)\chi_0 = 0\ .
\end{eqnarray}
Then, assuming $\chi_0\ne 0$, we arrive at
\begin{eqnarray}\label{HJ}
g^{\mu\nu}\partial_\mu S_{cl} \partial_\mu S_{cl} + m^2 c^2 = 0 \ ,
\end{eqnarray}
which is interpreted as the relativistic Hamilton-Jacobi equation. From here we can read off the relativistic Hamiltonian of a particle (for the details of the following procedure, see, e.g. \cite{Rovelli:2004tv})
\begin{eqnarray}\label{relham}
H = g^{\mu\nu} p_\mu p_\nu + m^2 c^2 \ ,
\end{eqnarray}
where we used the standard relation between the Hamilton's principal function, $S_{cl}$, and the canonical momentum:
\begin{eqnarray}\label{Sprelation}
p_\mu = \partial_\mu S_{cl} \ .
\end{eqnarray}
Of course, (\ref{relham}) is nothing but the usual relativistic dispersion relation. But the way it has been arrived at will prove to be useful for more general theories where the Diff symmetry will be broken and the dispersion relation will be not so obvious (see the section \ref{anyzotropicgeodesic}).

As usual for time reparametrization invariant theories, the relativistic Hamiltonian represents a constrain and the full dynamics of the system with respect to some affine parameter $\tau$ is given by the set of the equations:
\begin{eqnarray}\label{constdyn}
\left\{
  \begin{array}{l}
    H = 0 \\
    \dot{x}^\mu = N(\tau)\frac{\partial H}{\partial p_\mu} \\
    \dot{p}_\mu = - N(\tau)\frac{\partial H}{\partial x^\mu} \\
  \end{array}
\right. \ ,
\end{eqnarray}
which in our case becomes
\begin{eqnarray}\label{eom1}
\left\{
  \begin{array}{l}
    g^{\mu\nu} p_\mu p_\nu + m^2 c^2 = 0 \\
    \dot{x}^\mu = 2N(\tau)g^{\mu\nu} p_\nu \\
    \dot{p}_\mu = - N(\tau)\frac{\partial g^{\nu\lambda}}{\partial x^\mu} p_\nu p_\lambda \\
  \end{array}
\right. \ .
\end{eqnarray}
\textbf{Claim.} \textit{The set of equations (\ref{eom1}) is equivalent to (\ref{geodesic}).}\\
\begin{proof}
Combining the first and the second equations of (\ref{eom1}), we get
\begin{eqnarray}
\dot{x}^\mu p_\mu = -2 m^2 c^2 N(\tau) \ , \ \dot{x}^\mu \dot{x}_\mu = 2 N(\tau)\dot{x}^\mu p_\mu \ .
\end{eqnarray}
From here, we immediately obtain:
\begin{eqnarray}
m c N(\tau) = \frac{1}{2}\sqrt{-g_{\mu\nu}\dot{x}^\mu \dot{x}^\nu} \ .
\end{eqnarray}
Fix the gauge by $2N(\tau) = 1$, which essentially fixes $\tau$ to be proper time. Combining this with the second and the third equations of the system (\ref{eom1}) we get
\begin{eqnarray}
\ddot{x}^\mu & = & \frac{d}{d\tau}\left( g^{\mu\nu} p_\nu \right) = \partial_\lambda g^{\mu\nu} \dot{x}^\lambda p_\nu + g^{\mu\nu} \dot{p}_\nu = \partial_\lambda g^{\mu\nu} \dot{x}^\lambda p_\nu - \frac{1}{2} g^{\mu\nu}\frac{\partial g^{\rho\lambda}}{\partial x^\nu} p_\rho p_\lambda = \nonumber \\
& = & \partial_\lambda g^{\mu\nu} \dot{x}^\lambda g_{\nu\sigma}\dot{x}^\sigma - \frac{1}{2} g^{\mu\nu}\frac{\partial g^{\rho\lambda}}{\partial x^\nu}g_{\rho\sigma}\dot{x}^\sigma g_{\lambda\eta}\dot{x}^\eta = - \frac{1}{2}g^{\mu\nu}(\partial_\lambda g_{\nu\sigma} + \partial_\sigma g_{\nu\lambda} - \partial_\nu g_{\lambda\sigma})\dot{x}^\lambda \dot{x}^\sigma \equiv \nonumber \\
& \equiv & - \Gamma^\mu_{\lambda\sigma}\dot{x}^\lambda \dot{x}^\sigma \ .\nonumber
\end{eqnarray}
\end{proof}

This completes the demonstration that the physical motion of a test particle follows geodesics of the underlying geometry.

\subsection{Geodesics from the non-commutative geometry approach}\label{geodesicNC}

In this part, we would like to show how one can recover the geodesics for the standard case of Riemannian geometry. This is essentially a review of the well-known result, see e.g. \cite{GraciaBondia:2001tr},\cite{Martinetti:2001fq} (see also \cite{Martinetti:2012sc},\cite{D'Andrea:2013nda} for some applications), but we include it for the benefit of a reader because we will use the same approach in the next section for the case of a foliated space-time and we will see that in that more general case the present derivation can be adopted almost without any change. In our presentation, we will not try to be rigorous, but instead will present the steps essential to arrive at the result.

The starting point is some set of algebraic data that completely encodes the information about the Riemannian geometry\footnote{The fact that this procedure is given for the Riemannian geometry should not be a problem as we are assuming that the space is globally hyperbolic, as should be the case if we want to have a well-defined Cauchy problem \cite{Geroch:1970uw},  and eventually can pass to the Minkowski signature.}. Actually, we will look at a single aspect of this geometry, namely the definition of the shortest distance between two points (pure states in the algebraic languege). As we will see, it will be given exactly by the geodesic distance along a geodesic defined by (\ref{geodesic}). For the review and the complete treatment of non-commutative geometry, see e.g. \cite{GraciaBondia:2001tr},\cite{Connes:1994yd}.

The aforementioned set of algebraic data is called the spectral triple, $(\mathcal{A,H,D})$, and for the usual case of a Riemannian spin manifold $\mathcal{M}$ is given by

1) A $C^*$-algebra of the differentiable functions on $\mathcal{M}$, $\mathcal{A}=\mathcal{C}^\infty(\mathcal{M})$.

2) A Hilbert space of smooth square integrable sections of a spin bundle over $\mathcal{M}$, $\mathcal{H}=L_2(\mathcal{M,S})$. The scalar product in $\mathcal{H}$ is defined by $\langle\psi|\chi\rangle := \int_\mathcal{M} d^n x \sqrt{g}\bar{\psi}(x)\chi(x)$. (Compare the second term in (\ref{fullalaction}) and (\ref{action}).)

3) The usual Dirac operator (\ref{Dirac}), $\mathcal{D}=\mathrm{D}=\gamma^\mu \nabla^\omega_\mu$.

In terms of this data the distance between two points is \textit{defined} to be
\begin{eqnarray}\label{distance}
d(x,y) := \sup_{f\in\mathcal{C}^\infty(\mathcal{M})}\left\{ |f(x)-f(y)| : \|[\mathrm{D},f]\|\leq 1\right\} \ .
\end{eqnarray}
The norm used in (\ref{distance}) is the operator norm. As we will see below, we will need this norm only for an element of $\mathcal{C}^\infty(\mathcal{M})$. Then we can write for $f\in\mathcal{C}^\infty(\mathcal{M})$
\begin{eqnarray}\label{norm}
\|f\| := \sup_{\psi\in\mathcal{H}}\left[ \frac{\int_\mathcal{M} d^n x \sqrt{g}f^*\bar{\psi}(x)f\psi(x)}{\int_\mathcal{M} d^n x \sqrt{g}\bar{\psi}(x)\psi(x)}\right]^{1/2}\equiv  \sup_{x\in\mathcal{M}}|f(x)|\ .
\end{eqnarray}
(The last step in (\ref{norm}) is pretty trivial: Clearly the left hand side is less or equal then the right hand side. Then show that exists a spinor as close as needed to the one proportional to the ``square root'' of a delta-function with the support at the point where the supremum is reached.)

We need to find $\|[\mathrm{D},f]\|$. This is done as follows (using (\ref{norm}), the definition of gamma-matrices from the ppendix and the $C^*$-algebra property, $\|A\|^2 = \|A^* A\|$ for any element $A$ in this algebra. For the general properties of $C^*$-algebras and some of their physical applications, see e.g. \cite{Landsman:1998zs})
\begin{eqnarray}\label{normD}
\|[\mathrm{D},f]\|^2 &=& \|\gamma^\mu\partial_\mu f \|^2 = \|\gamma^\mu\partial_\mu f \gamma^\nu\partial_\nu f\| = \|g^{\mu\nu} \partial_\mu f \partial_\nu f\| = \sup_{x\in\mathcal{M}}|g^{\mu\nu} \partial_\mu f \partial_\nu f| = \nonumber\\
&=& \sup_{x\in\mathcal{M}}\|\mathrm{grad} f\|_x^2 := \|\mathrm{grad} f\|^2_\infty \ .
\end{eqnarray}
Here we used the so called musical isometric isomorphism between 1-forms and vectors: $(\mathrm{grad} f)^\mu = g^{\mu\nu} \partial_\nu f$. And the norm $\|\cdot\|_x$ is the norm on the tangent space $T_x\mathcal{M}$ defined by $g^{\mu\nu}$. So we can rewrite (\ref{distance}) in the following form
\begin{eqnarray}\label{distance1}
d(x,y) := \sup_{f\in\mathcal{C}^\infty(\mathcal{M})}\left\{ |f(x)-f(y)| : \|\mathrm{grad} f\|_\infty\leq 1\right\} \ .
\end{eqnarray}

What we want to do now is to compare (\ref{distance1}) with the usual geodesic distance on a Riemannian manifold:
\begin{eqnarray}\label{distanceR}
d_R(x,y) := \inf_{\gamma}\left\{ length (\gamma) :\ \gamma : [0,1] \rightarrow\mathcal{M}, \gamma(0)=x,\gamma(1)=y\right\} \ ,
\end{eqnarray}
where $length (\gamma)$ is defined in the usual way with the help of the metric as in (\ref{length}). For any $f\in\mathcal{C}^\infty(\mathcal{M})$ and any $\gamma$ we can write
\begin{eqnarray}\label{fyfx}
f(y) - f(x) = f(\gamma(1)) - f(\gamma(0)) = \int\limits^1_0 \frac{d}{d\tau}[f(\gamma(\tau)]d\tau = \int\limits^1_0 [g_{\mu\nu}(\mathrm{grad}f)^\mu \dot{\gamma}^\nu]_{\gamma(\tau)}d\tau\ .
\end{eqnarray}
Then using the Cauchy-Schwarz-Bunyakovsky inequality we arrive at the following estimate
\begin{eqnarray}\label{absfyfx}
|f(y) - f(x)| \le \int\limits^1_0 \|\mathrm{grad}f\|_{\gamma(\tau)} \|\dot{\gamma}\|_{\gamma(\tau)}d\tau \le \|\mathrm{grad}f\|_\infty\int\limits^1_0 \|\dot{\gamma}\|_{\gamma(\tau)}d\tau \equiv \|\mathrm{grad}f\|_\infty length (\gamma)\ .
\end{eqnarray}
Because (\ref{absfyfx}) is valid for any $f$ and any $\gamma$, using (\ref{distance1}) and (\ref{distanceR}) we arrive at the following inequality:
\begin{eqnarray}\label{bound}
d(x,y)\le d_R (x,y)\ .
\end{eqnarray}
It is easy to demonstrate that this inequality is saturated. Really, choose the function $f$ to be
\begin{eqnarray}\label{fx}
f_x (y) :=  d_R (x,y)\ .
\end{eqnarray}
It is trivial fact that $\|\mathrm{grad}f_x\|_\infty = 1$ (actually $\mathrm{grad}f_x$ is nothing but a unit 4-velosity). Observing that $|f_x(y) - f_x(x)| = d_R (x,y)$ completes the proof that
\begin{eqnarray}\label{ddR}
d(x,y) = d_R (x,y)\ .
\end{eqnarray}

So, we can see that the same geodesics follow from two completely different approaches. The first one, described in the section \ref{HJstandard} is based on the dynamics of a physical system, while the second outlined in this section is purely geometric (or rather algebraic) for which we do not need the presence of matter. Of course, the trick is that the same object, the Dirac operator (\ref{Dirac}), crucially enters both constructions. In what follows, we will address the following question: Does this agreement hold if the physical theory as well as the geometry are deformed but still share the same (deformed) Dirac operator?

\section{Geodesics for a foliated space-time}\label{anyzotropicgeodesic}

Now we want to generalize both approaches described in the previous section to the case of a more general coupling of matter to gravity. Namely, in the spirit of the noncommutative geometry and spectral action principle, we would like to keep the same form of the action (\ref{action}) but use more general Dirac operator. This would still correspond to the minimal coupling between matter and, now generalized, geometry.

In the appendix \ref{split} we have found the 3+1 decomposition of the standard Dirac operator (\ref{3plus1})
\begin{eqnarray}\label{3plus11}
\mathrm{D} = \gamma^0 D_n + {}^{(3)}\mathrm{D} - \frac{1}{2}\gamma^0 K - \frac{1}{2}\gamma^\alpha \frac{\partial_\alpha N}{N}\ .
\end{eqnarray}
This operator respects the full Diff symmetry, meaning that all the coefficients in front of each of the four terms in (\ref{3plus11}) should be exactly as they are if we insist on Diff covariance. On the other hand, if we only interested in FPDiff, it can be shown that each term in (\ref{3plus11}) is \textit{separately} FPDiff covariant. This leads to the natural generalization of $\mathrm{D}$
\begin{eqnarray}\label{Diracgen}
\mathbb{D}= \gamma^0 D_n + c_1{}^{(3)}\mathrm{D} + c_2\gamma^0 K + c_3\gamma^\alpha a_\alpha\ ,
\end{eqnarray}
where we introduced $a_\alpha := \frac{\partial_\alpha N}{N}$. Actually motivated by the HL-type theories \cite{Horava:2009uw},\cite{Sotiriou:2010wn}, we could insist on the anisotropic scaling in UV with $z=3$, then (\ref{Diracgen}) is not the most general form of the generalized Dirac operator - we still can add higher space derivative terms. Taking into account that $[D_n] = [K] =z$ and $[{}^{(3)}\mathrm{D}] = [a_\alpha] =1$,\footnote{This is not hard to show looking at the metric (\ref{ADM}) that we have the following scaling dimensions: $[N]=[h_{\mu\nu}]=0$, $[N^\alpha]=z-1$. Then the scaling dimensions for the terms in $\mathrm{D}$ follow.} the resulting form of the operator would be
\begin{eqnarray}\label{DiracgenUV}
\mathbb{D}_{UV}= \gamma^0 D_n + c\gamma^0 K + \sum\limits_{n+m\le 3} c_{nm}{}^{(3)}\mathrm{D}^n (\gamma^\alpha a_\alpha)^m\ .
\end{eqnarray}
This expression has 11 free parameters (mass being one of them for $n=m=0$). This should be compared with the number of free parameters in the general Horava-Lifshitz gravity, which is of the order of 100 \cite{Blas:2010hb}. Postponing the analysis of the complete generalized operator (\ref{DiracgenUV}) for the future, here we will be interested in its IR version (\ref{Diracgen}), i.e. when we can neglect the terms with higher derivatives.

We want to study the effect produced by this generalization on the physical motion of matter particles. As it was stressed before, now we do not have a conserved energy-momentum tensor, so there is no reason to expect that the physical motion will be geodesic one for the underlying metric. Nevertheless we will see that we still can say that the physical motion of a test particle is along geodesics but for some new geometry. For this, we will analyze the question of the physical motion of a test particle and the definition of geodesics using both approaches considered above.

\subsection{Physical motion from the Hamilton-Jacobi eqation}\label{physmotionNC}

Here we apply the method from the section \ref{HJstandard} for the case of the action (\ref{action}) and the Dirac operator $\mathrm{D}_m := \mathbb{D} - \frac{mc}{\hbar}$. The equation of motion take exactly the same form as in the standard case
\begin{eqnarray}\label{Diracequationgen}
\left(\mathbb{D} - \frac{mc}{\hbar}\right)\psi =0 \ .
\end{eqnarray}
Proceeding as before, writing $\psi$ as in (\ref{quasiclassic1}), we can re-write (\ref{Diracequationgen}) as
\begin{eqnarray}\label{Diracequationgen1}
\mathbb{D} \chi - \frac{i}{\hbar}\gamma^0 \chi n^\mu \partial_\mu S + \frac{i}{\hbar}c_1\gamma^\alpha \chi \partial_\alpha S - \frac{mc}{\hbar}\chi =0 \ ,
\end{eqnarray}
and after using the same quasiclassical expansion of $\chi$ and $S$ as in (\ref{quasiclassicexpansion}) we arrive at the leading $\frac{1}{\hbar}$ order at the following equation
\begin{eqnarray}\label{Diracequasiclassic}
i\gamma^0 \chi_0 n^\mu \partial_\mu S_{cl} + ic_1\gamma^\alpha \chi_0 \partial_\alpha S_{cl} - mc\chi_0 =0 \ .
\end{eqnarray}
There are two important points about (\ref{Diracequasiclassic}): 1) Compared to (\ref{quasiclassicfermion1}), it depends on the parameter $c_1$ and will reduce to (\ref{quasiclassicfermion1}) only if $c_1 = 1$; 2) The other two parameters present in (\ref{Diracgen}), $c_2$ and $c_3$, do not enter (\ref{Diracequasiclassic}) and, as the consequence, cannot affect the classical physical motion in this model. The second point is quite interesting and not completely expected: varying $c_2$ and $c_3$ can lead to an arbitrary large braking of Diff symmetry and yet it will not manifest itself in any gravitational experiment on a test particle motion. Of course, as we will discuss later, these parameters should be constrained from the other point of view in the framework of the spectral action.

To complete our study of the physical motion of a test particle, we need to exclude $\chi_0$ from (\ref{Diracequasiclassic}) as it was done for (\ref{quasiclassicfermion1}). To do so, we re-write (\ref{Diracequasiclassic}) in the form
\begin{eqnarray}\label{chi0}
\chi_0 =\frac{i}{mc}(\gamma^0 \chi_0 n^\mu \partial_\mu S_{cl} + c_1\gamma^\alpha \chi_0 \partial_\alpha S_{cl} )
\end{eqnarray}
and plug it back into (\ref{Diracequasiclassic}) arriving at the following equation
\begin{eqnarray}
(-n^\mu n^\nu \partial_\mu S_{cl}\partial_\nu S_{cl} + c_1^2 \gamma^\alpha \gamma^\beta \partial_\alpha S_{cl}\partial_\beta S_{cl} + m^2 c^2)\chi_0 = 0 \ .
\end{eqnarray}
Using $\{\gamma^\alpha , \gamma^\beta\} = e_i^{ \alpha} e_j^{ \beta} \{\gamma^i , \gamma^j\}=2h^{\alpha\beta}$ (see (\ref{projector1})) and again assuming $\chi_0 \ne 0$ we arrive at the deformed analog of the Hamilton-Jacobi equation (\ref{HJ})
\begin{eqnarray}\label{HJdefermed}
(-n^\mu n^\nu + c_1^2 h^{\mu\nu}) \partial_\mu S_{cl}\partial_\nu S_{cl} + m^2 c^2 = 0 \ .
\end{eqnarray}
We see that the whole effect has been reduced to the re-scaling of the space metric $h^{\mu\nu}$. If we formally introduce a new metric $\tilde{g}$ by (see (\ref{projector}))
\begin{eqnarray}\label{newmetric}
\tilde{g}_{\mu\nu} = -n_\mu n_\nu + \frac{1}{c_1^2} h_{\mu\nu}  \ ,
\end{eqnarray}
then in terms of this metric, (\ref{HJdefermed}) takes exactly the same form as in the standard case (\ref{HJ})
\begin{eqnarray}\label{HJtil}
\tilde{g}^{\mu\nu}\partial_\mu S_{cl} \partial_\mu S_{cl} + m^2 c^2 = 0 \ ,
\end{eqnarray}
which immediately allows to repeat the whole analysis of the section \ref{HJstandard} that follows the equation (\ref{HJ}). Doing so, the result for the physical motion will take exactly the same form as for the standard geodesics (\ref{geodesic}) but for the modified (but still Riemannian!) geometry:
\begin{eqnarray}\label{geodesicdeformed}
\frac{d^2 x^\mu}{d \tau^2}+\tilde{\Gamma}^\mu_{\nu\lambda}\frac{d x^\nu}{d \tau}\frac{d x^\lambda}{d \tau} = 0 \ ,
\end{eqnarray}
where $\tilde{\Gamma}^\mu_{\nu\lambda}$ are the Christoffel symbols calculated for the new metric $\tilde{g}_{\mu\nu}$ and $\tau$ is the modified proper time defined by $d\tau = \sqrt {- \tilde{g}_{\mu\nu}dx^\mu dx^\nu }$.

\subsection{Generalized geodesics from the non-commutative geometry approach}

Here we would like to see what will result in application of the approach of the section \ref{geodesicNC} if we use it for the generalized geometry defined by a new spectral triple. The only difference of this new spectral triple from the standard one used in the section \ref{geodesicNC} is that instead of the standard Dirac operator (\ref{Dirac}) we will use its deformed version (\ref{Diracgen}). The point of asking such a question is the following: if we will be able to show that the equation (\ref{geodesicdeformed}) defines the shortest distance paths in the sense of the definition of distance (\ref{distance}), as was the case for the standard Dirac operator (see (\ref{ddR})), we will conclude that the physical motion is still a geodesic one but in the generalized geometry defined by the deformed spectral triple. If this is true, then the metric (\ref{ADM}) is not the physical one but plays some auxiliary role and the physical one is $\tilde{g}_{\mu\nu}$ (at least as far as the metric properties of the generalized geometry are concerned). As we now show, this is indeed so.

In fact we almost do not have to do any calculations. The only way the Dirac operator enters the definition of distance (\ref{distance}) is through $[\mathbb{D},f]$. It is trivial to see that again only $c_1$ constant matters:
\begin{eqnarray}\label{Dfcomm}
[\mathbb{D},f] = \gamma^0 n^\mu \partial_\mu f + c_1 \gamma^\alpha \partial_\alpha f \ ,
\end{eqnarray}
where $f(x)\in\mathcal{C}^\infty (\mathcal{M})$. As we stressed above, for the case of $c_1 = 1$ we recover the standard case of the section (\ref{geodesicNC}) (again, $c_2$ and $c_3$ do not matter!). We can easily bring (\ref{Dfcomm}) to the form that will look exactly as in the standard case. This is achieved by the appropriate re-scaling of the space tetrads (recall that $\gamma^\alpha = e_i^{\ \alpha}\gamma^i$, see appendix):
\begin{eqnarray}\label{rescaling}
\tilde{e}_i^{\ \alpha} := c_1 e_i^{\ \alpha} \ .
\end{eqnarray}
In view of (\ref{projector1}), it is clear that the re-scaling (\ref{rescaling}) leads to exactly the same modified metric $\tilde{g}$ (\ref{newmetric}). Then we again can just repeat the rest of the calculation from the section (\ref{geodesicNC}) to conclude that the equation for the physical motion (\ref{geodesicdeformed}) indeed defines the path of the shortest distance in the sense of the modified (but still Riemannian) geometry.\footnote{In fact, from the point of view of non-commutative geometry approach the fact that the geodesics defined by the spectral triple based on the Dirac operator (\ref{Diracgen}) are given by the usual geodesics in the usual Riemannian geometry, i.e. by (\ref{geodesicdeformed}), is expectable. The point is that this spectral triple satisfies all the requirements of the so-called \textit{reconstruction theorem} \cite{Connes:2008vs}. This theorem, in its turn, insures that the resulting geometry defined by this spectral triple will be some usual Riemannian geometry. What is non-trivial is that the physical motion deduced from completely different principles in section \ref{physmotionNC} still follows these geodesics. Using this observation we can see that we should not expect such simplification as existence of some usual Riemannian geometry for the spectral triple defined by the fully deformed Dirac operator (\ref{DiracgenUV}). This is because in this case at least one of the conditions (the so-called \textit{the first order condition}) of the reconstruction theorem is violated. This violation happens exactly due to the presence the higher order derivatives in (\ref{DiracgenUV}). I am grateful to Dmitri Vassilevich for bringing this to my attention.}

\section{Discussion and conclusion}

In this paper, we have addressed the question of the correspondence between the physical motion of a test matter particle in a given geometry of space-time (due to the not necessarily Diff-invariant interaction with this geometry) and the notion of the geodesics for the same geometry. While in the standard case of pseudo-Riemannian geometry and minimally coupled matter, i.e. in General Relativity, can be shown that the physical motion \textit{is} the geodesic one, usually it is done relying on the covariantly conserved energy-momentum tensor \cite{Dixon:1970zza},\cite{Dixon:1970zz},\cite{Hawking:1973uf}. Motivated by the HL-type theories of gravity where such tensor does not exist in general, we develop the alternative derivation of this fact based on the quasiclassical analysis of field equations. At the same time, motivated by our earlier work \cite{Pinzul:2010ct} as well as work in progress \cite{LMP}, which hint on the possibility of the existence of some underlying generalized geometry in HL-type theories, we review the purely algebraic way of deriving geodesics that relies on a Dirac operator.

After demonstrating our approach for the standard case, we pass to study the matter minimally coupled to the foliated geometry. In this work, we study only the minimal deformation of the standard case, restricting ourselves to the first order Dirac operator (\ref{Diracgen}), which should be thought of as working in the IR regime. But even in this case the full Diff symmetry is broken to FPDiff in matter sector. As the consequence, there is no covariantly conserved energy-momentum tensor and we have to rely on our approach. The main results of our consideration are following.

Firstly, we do find that the physical motion will deviate from the standard geodesics defined for the geometry (\ref{ADM}). While this is not unexpected, the explicit form of this deviation has not been known to our knowledge.

This brings the second, less expected, result: the deviation from the geodesic motion depends only on one parameter of the 3-dim parameter space. This potentially can have serious experimental consequences (see also below).

Thirdly, we establish that even in this deformed case, one still can speak of the geodesic motion. The difference with the standard case is in the geometry: now these geodesics are defined for the geometry different from the one we started with. Due to the first order of the deformed Dirac operator (\ref{Diracgen}), this modified geometry still can be written in terms of a metric (though different from the original). We expect that this will not longer be true for the general Dirac operator $\mathbb{D}_{UV}$ (\ref{DiracgenUV}). This question is being studied.

Also we would like to mention that the 3+1 decomposition of the Dirac operator found in this work (see the appendix) should prove useful in the future studies of matter in HL-type theories. This brings the question of what is next to do.

The most urgent question is ``Can we see the deviations described by (\ref{geodesicdeformed})?''. To answer this question we need to know the metric (\ref{ADM}), which, in this IR regime, is the solution of the gravitational equations of motion following from the most general IR FPDiff invariant action \cite{Barausse:2013nwa}
\begin{eqnarray}\label{IRaction}
S_{IR} = \frac{1}{16\pi G} \int dt\, d^3 x N\sqrt(h) (K_{\alpha\beta}K^{\alpha\beta}-\lambda K^2 + \xi {}^{(3)}R + \eta a_\alpha a^\alpha)\ .
\end{eqnarray}
As we can see this action has its own 3-dim parameter space $(\lambda , \xi , \eta)$. If these parameters are independent of $(c_1 , c_2 , c_3)$ from (\ref{Diracgen}) then the question of the possible deviations is not that interesting: we have too many independent parameters to play with. The situation changes drastically if there is some relation between these two sets of the parameters. This is the case if one insists that (\ref{IRaction}) comes from the spectral action \cite{Chamseddine:1996zu} defined by \textit{the same} Dirac operator used in the matter sector (see the first term in (\ref{fullalaction})):
\begin{eqnarray}\label{spectralaction}
S_{IR} = \Tr f\left(\frac{\mathbb{D}^2}{\Lambda^2}\right)\ ,
\end{eqnarray}
where $f$ is some cut-off function and $\Lambda$ is a characteristic scale (could be the same scale as in HL model). Then it is clear that $(\lambda , \xi , \eta)$ are not independent anymore but rather are some functions of $c_i$ (and possibly $\Lambda$). This is very interesting because now two types of the corrections (or deviations), one coming from the fact that the metric itself is \textit{not} the same as in Einstein theory (see e.g. \cite{Barausse:2013nwa} for the example of Schwarzschild metric), and the second being found in this paper, become dependent. This point is very important because this will affect bounds on the parameter space coming from the observations of test particles, e.g. Solar system experiments. In the extreme case both deviations could exactly cancel and the some other experiment will be needed to tell between HL-type theory an GR. And what about the terms in (\ref{Diracgen})proportional $c_2$ and $c_3$? Though these terms do not affect the physical motion in a given geometrical background, they will have effect in the form of the Lorentz breaking. Nevertheless, it is crucial to note that these terms are proportional some geometrical quantities, $K$ and $a_\alpha$, which a re zero for the flat case or small for weak gravity. These effects should be studied in the general framework of Lorentz violating theories \cite{Colladay:1998fq}. Such analysis should lead to stringent bounds on $c_2$ and $c_3$, which in its turn will constrain the parameters of the gravitational part (\ref{IRaction}) of the spectral action. These important questions are currently under study \cite{LMP}.

The more ambitious (and more important) problem is recovery of the full HL-type theory coupled to matter starting with the fully deformed Dirac operator $\mathbb{D}_{UV}$ (\ref{DiracgenUV}). Along with providing the natural (minimal) way of the coupling of matter to the HL gravity (the second term in (\ref{fullalaction})), the approach based on the spectral action will enormously reduce the number of free parameters. As was mentioned in the section \ref{anyzotropicgeodesic}, the most general Dirac operator respecting FPDiff symmetry contains only 11 parameters. Using this operator in the spectral action (\ref{spectralaction}) will produce a HL-type theory that also will depend only on these parameters (and some scale $\Lambda$). This is in great contrast with the present situation \cite{Blas:2010hb}. Unfortunately the problem of constructing the spectral action based on $\mathbb{D}_{UV}$ is not the easy one both technically and conceptually. The first steps in this direction were undertaken in \cite{Mamiya:2013wqa} where the heat kernel for the flat anisotropic foliated space-time was calculated. The further work in this direction is in progress and the developments will be reported elsewhere.

\section*{Acknowledgements}

The author would like to thank Daniel Lopes and Arthur Mamiya for the insightful discussions during different stages of this project and Dmitri Vassilevich for useful comments. The work was done under the partial support of CNPq under grant no.306068/2012-5.

\appendix
\section{Notations and conventions}

Here we fix the notations and conventions used in the main text as well as establish some formulas used in the further calculations.

\textit{Coordinate system.} Our space-time $\mathcal{M}$ has the structure of a foliation. Because in this work this structure is considered to be fundamental, it is natural to adopt the following coordinates:
\begin{eqnarray}\label{coord}
x^\mu = (t, \vec{x})\ ,
\end{eqnarray}
where $t=const$ defines a leaf of the foliation $\Sigma_t$, while $\vec{x}$ are the coordinates on $\Sigma_t$.

\textit{Metric.} We are using the metric with the signature $\mathrm{sign}\ g = 2$, i.e. a time-like vector $n^\mu$ has a negative length.\footnote{Though we do specify the metric to be a pseudo-Riemannian one, the whole consideration exactly goes through (modulo some sign changes) for any signature.} In the coordinates (\ref{coord}), the metric takes the usual ADM form \cite{Arnowitt:1962hi}
\begin{eqnarray}\label{ADM}
d s^2 = - (Ndt)^2 + h_{\alpha\beta} (dx^\alpha + N^\alpha dt)(dx^\beta + N^\beta dt)\ ,
\end{eqnarray}
where $N$ is the laps function and $N^\alpha$ is the shift vector. Throughout the paper we are using the following system of indices:

The Greek letters from the middle of the alphabet, $\mu,\ \nu,\ \ldots $ are used to denote the curved coordinates (\ref{coord}) and take values 0, 1, 2 and 3.

The Greek letters from the beginning of the alphabet, $\alpha,\ \beta,\ \ldots $ are used to denote the space part of the curved coordinates (\ref{coord}), i.e. $\vec{x}$, and take values 1, 2 and 3.

The Latin letters from the beginning of the alphabet, $a,\ b,\ \ldots $ are used to denote the flat coordinates of 4d Minkowski space and take values 0, 1, 2 and 3.

The Latin letters from the middle of the alphabet, $i,\ j,\ \ldots $ are used to denote the flat coordinates of the space part of 4d Minkowski space and take values 1, 2 and 3.

\textit{Tetrads, Second fundamental form.} We partially fix the local $SO(3,1)$ invariance (which is natural to do keeping in mind the fundamental meaning of the foliation) and choose the time-like tetrad to be equal to the vector normal to $\Sigma_t$, i.e.
\begin{eqnarray}\label{zerotetrad}
e_0^{\ \mu} = n^\mu\ ,
\end{eqnarray}
where $n^\mu$ is the vector dual to the 1-form $n=N dt$. Clearly, this vector is normal to the hypersurface $t=const$, and using (\ref{ADM}) we see that
\begin{eqnarray}\label{normal}
n_\mu=(N,0,0,0)\ , \ n^\mu = \left(-\frac{1}{N},\frac{N^\alpha}{N}\right)\ \mathrm{and} \ n^\mu n_\mu = -1 \ .
\end{eqnarray}
Then the rest of the tetrads will belong to the space tangent to $\Sigma_t$, $e_i^{\ \mu}\in \mathrm{T}\Sigma_t$. As usual, we can introduce the projector on $\mathrm{T}\Sigma_t$
\begin{eqnarray}\label{projector}
h_{\mu\nu} = g_{\mu\nu} + n_\mu n_\nu \ .
\end{eqnarray}
The fact that $h_{\mu\nu}$ projects any vector from $\mathrm{T}\mathcal{M}$ to a vector in $\mathrm{T}\Sigma_t$, immediately follows from that a) $h_{\mu\nu} n^\mu = 0$ and b) $h_{\mu}^{\ \rho} h_{\rho\nu} = h_{\mu\nu}$. Using this, one can easily see that $e_i^{\ \mu}$ are left invariant by $h_{\mu\nu}$, $h^{\mu}_{\ \nu} e_i^{\ \nu} = e_i^{\ \mu}$. Also, combining (\ref{zerotetrad}), (\ref{projector}) and $e_{a\mu} e^{a}_{\ \nu}=g_{\mu\nu}$, one can establish that
\begin{eqnarray}\label{projector1}
h_{\mu\nu} = e_{i\mu} e^{i}_{\ \nu} \ .
\end{eqnarray}
Also note that due to $n_\mu e_i^{\ \mu} = 0$, we have
\begin{eqnarray}\label{spacetetrad}
e_i^{\ \mu} = (0, e_i^{\ \alpha}) \ .
\end{eqnarray}
This will be used in the second part of the appendix.

Using the normal vector $n^\mu$ and the projector $h_{\mu\nu}$ we define in the standard way a second fundamental form, or the extrinsic curvature, which measures how the leaves of the foliaton are ``bended'' in the ambient space-time
\begin{eqnarray}\label{externalcurvature}
K_{\mu\nu} = -h_\mu^{\ \rho}\nabla_\rho n_\nu\ \ .
\end{eqnarray}

\textit{Covariant spin derivative.} As usual, to work with fermions in a curved space-time, we need to introduce an appropriate derivative, see, e.g. \cite{Nakahara:1990th}:
\begin{eqnarray}\label{spincovder}
\nabla^\omega_\mu = \partial_\mu + \omega_\mu \ ,
\end{eqnarray}
where $\omega_\mu = \frac{1}{4}\omega_{\mu ab}\gamma^{ab}$ is a spin connection and $\gamma^{ab}:= \frac{1}{2}[\gamma^{a},\gamma^{b}]$ are the generators of $SO(3,1)$. Here $\gamma^{a}$ are the usual flat gamma matrices, i.e. $\{\gamma^{a},\gamma^{b}\} = 2\eta^{ab}$. The condition that covariant derivative is compatible with metric is translated into the full (i.e. with respect to both space-time and flat indices) covariant constancy of the tetrad:
\begin{eqnarray}
\tilde{\nabla}_\mu e_{a\nu} \equiv \partial_\mu e_{a\nu} + \omega_{\mu a}^{\ \ b}e_{b\nu} - \Gamma^{\rho}_{\mu\nu}e_{a\rho} = 0\ .
\end{eqnarray}
From here, it is easy to find the expression for $\omega_{\mu ab}$
\begin{eqnarray}\label{spinconnection}
\omega_{\mu ab} = e_{a\nu}\partial_\mu e_{b}^{\ \nu} + \Gamma^{\rho}_{\mu\nu}e_{a\rho}e_{b}^{\ \nu}\equiv e_{a\nu}\nabla_\mu e_{b}^{\ \nu}\ ,
\end{eqnarray}
where now $\nabla_\mu$ is the usual space-time covariant derivative, i.e. the one acting on space-time indices only.

\textit{Dirac operator.} We define the Dirac operator in the standard way by
\begin{eqnarray}\label{Dirac}
\mathrm{D} = \gamma^\mu \nabla^\omega_\mu \ ,
\end{eqnarray}
where $\gamma^\mu := e_{a}^{\ \mu}\gamma^a$ are the curved gamma matrices, i.e. $\{\gamma^\mu,\gamma^\nu\} = 2g^{\mu\nu}$ (which is trivial by $e_{a\mu} e^{a}_{\ \nu}=g_{\mu\nu}$).

\section{3+1 split of the Dirac operator}\label{split}

Here our goal is to decompose the Dirac operator (\ref{Dirac}) in terms of the ADM variables, i.e. in terms of 3d metric $h_{\mu\nu}$ and the second fundamental form $K_{\mu\nu}$.\footnote{The author greatly benefited from the discussions on 3+1 decomposition with Arthur Mamiya.} Towards this end we write using (\ref{projector})
\begin{eqnarray}\label{Diracdecomposed}
\mathrm{D} = g^{\mu\nu}\gamma_\mu \nabla^\omega_\nu = (h^{\mu\nu} - n^\mu n^\nu) \gamma_\mu \nabla^\omega_\nu\ .
\end{eqnarray}
So, we need to analyze two terms: 1) $h^{\mu\nu} \gamma_\mu \nabla^\omega_\nu$ and 2) $n^\mu n^\nu \gamma_\mu \nabla^\omega_\nu $.

1) $h^{\mu\nu} \gamma_\mu \nabla^\omega_\nu$

Using (\ref{zerotetrad}) and (\ref{spacetetrad}), we have
\begin{eqnarray}\label{11}
h^{\mu\nu} \gamma_\mu \nabla^\omega_\nu = h^{\mu\nu} e_{a\mu}\gamma^a \nabla^\omega_\nu = e_{i}^{\ \mu}\gamma^i \nabla^\omega_\mu = \gamma^\alpha \partial_\alpha +\gamma^i e_{i}^{\ \mu}\omega_\mu\ .
\end{eqnarray}
While the first term already contains just space derivatives, i.e. has already been projected to the hypersurface $t=const$, the second one requires more care. Using (\ref{zerotetrad}), (\ref{externalcurvature}) and (\ref{spinconnection}), we have
\begin{eqnarray}\label{12}
e_{i}^{\ \mu}\omega_\mu &=& \frac{1}{8}e_{i}^{\ \mu}\omega_{\mu ab}[\gamma^a , \gamma^b] = \frac{1}{8}e_{i}^{\ \mu}e_{a\nu}\nabla_\mu e_{b}^{\ \nu}[\gamma^a , \gamma^b] = \nonumber\\
&=& -\frac{1}{2} e_{i}^{\ \mu}e_{j\nu}\nabla_\mu e_{0}^{\ \nu}\gamma^0 \gamma^j + \frac{1}{4} e_{i}^{\ \mu}e_{j\nu}\nabla_\mu e_{k}^{\ \nu}\gamma^j \gamma^k = \nonumber\\
&=& -\frac{1}{2} e_{i}^{\ \mu}e_{j\nu}h_\mu^{\ \rho}\nabla_\rho n^{\nu}\gamma^0 \gamma^j + \frac{1}{4} e_{i}^{\ \mu}e_{j\nu}h_\mu^{\ \rho}\nabla_\rho e_{k}^{\ \nu}\gamma^j \gamma^k = \nonumber\\
&=& \frac{1}{2} e_{i}^{\ \mu}e_{j}^{\ \nu}K_{\mu\nu}\gamma^0 \gamma^j + e_{i}^{\ \alpha} {}^{(3)}\omega_\alpha \ .
\end{eqnarray}
Here ${}^{(3)}\omega_\alpha = \frac{1}{4}e_{j\beta}{}^{(3)}\nabla_\alpha e_{k}^{\ \beta}\gamma^j \gamma^k$ is the 3d spin connection, which trivially follows from the definition and the projective properties of $h_{\mu\nu}$. Combining (\ref{11}) and (\ref{12}) we get
\begin{eqnarray}\label{13}
h^{\mu\nu} \gamma_\mu \nabla^\omega_\nu = {}^{(3)}\mathrm{D} + \frac{1}{2} e_{i}^{\ \mu}e_{j}^{\ \nu}K_{\mu\nu}\gamma^i\gamma^0 \gamma^j \equiv {}^{(3)}\mathrm{D} - \frac{1}{2}\gamma^0 K\ ,
\end{eqnarray}
where we used that $e_{i}^{\ \mu}e_{j}^{\ \nu}K_{\mu\nu}\gamma^i \gamma^j=e_{i}^{\ \mu}e_{j}^{\ \nu}K_{\mu\nu}\delta^{ij}=h^{\mu\nu}K_{\mu\nu}=K$, see (\ref{projector1}) and introduced the 3d Dirac operator ${}^{(3)}\mathrm{D}$. Note, here and everywhere else $\gamma^0$ is a \textit{flat} gamma matrix.

2) $n^\mu n^\nu \gamma_\mu \nabla^\omega_\nu $

Using (\ref{zerotetrad}), we have
\begin{eqnarray}\label{21}
n^\mu n^\nu \gamma_\mu \nabla^\omega_\nu = n^\mu n^\nu e_{a\mu}\gamma^a \nabla^\omega_\nu = -\eta_{0a}n^\nu \gamma^a \nabla^\omega_\nu = \gamma^0 n^\nu \nabla^\omega_\nu =: \gamma^0 (\partial_n + n^\mu \omega_\mu)\ ,
\end{eqnarray}
where we have defined the derivative along the normal, $\partial_n := n^\mu \partial_\mu$, which for the suitable coordinates with the zero shift vector becomes just, up to a factor, a time derivative: $\partial_n = -\frac{1}{N} \partial_t$.

We still have to decompose $n^\mu \omega_\mu$:
\begin{eqnarray}\label{22}
n^\mu \omega_\mu = \frac{1}{8} n^\mu e_{a\nu}\nabla_\mu e_{b}^{\ \nu}[\gamma^a , \gamma^b] = \frac{1}{4} n^\mu n_\nu\nabla_\mu e_{i}^{\ \nu}[\gamma^0 , \gamma^i] + \frac{1}{8} n^\mu e_{i\nu}\nabla_\mu e_{j}^{\ \nu}[\gamma^i , \gamma^j] \ .
\end{eqnarray}
To deal with the first term in (\ref{22}) we will explicitly use the coordinate form of the normal vector (\ref{normal}) and the space tetrad (\ref{spacetetrad}):
\begin{eqnarray}\label{23}
\frac{1}{4} n^\mu n_\nu\nabla_\mu e_{i}^{\ \nu}[\gamma^0 , \gamma^i] &=& \frac{1}{2} n^\mu n_\nu (\partial_\mu e_{i}^{\ \nu} + \Gamma^{\nu}_{\mu\rho}e_{i}^{\ \rho})\gamma^0 \gamma^i = \frac{1}{2} \Gamma^{\nu}_{\mu\rho}e_{i}^{\ \rho}n^\mu n_\nu\gamma^0 \gamma^i = \nonumber\\
&=&\frac{1}{4}(g_{\nu\mu , \rho} + g_{\nu\rho ,\mu} - g_{\mu\rho ,\nu}) e_{i}^{\ \rho}n^\mu n^\nu\gamma^0 \gamma^i = \frac{1}{4}g_{\nu\mu , \rho} e_{i}^{\ \rho}n^\mu n^\nu\gamma^0 \gamma^i = \nonumber\\
&=& \frac{1}{4}(h_{\nu\mu , \rho} - n_{\nu , \rho} n_{\mu} - n_{\nu} n_{\mu , \rho}) e_{i}^{\ \rho}n^\mu n^\nu\gamma^0 \gamma^i = \frac{1}{2} e_{i}^{\ \rho}n^\mu n_{\mu , \rho}\gamma^0 \gamma^i = \nonumber\\
&=& - \frac{1}{2}\gamma^0 \gamma^i e_{i}^{\ \alpha}\frac{\partial_\alpha N}{N} \ ,
\end{eqnarray}
where we also used the fact that $h_{\nu\mu , \rho}n^\mu n^\nu = 0$, which immediately follows from $h_{\nu\mu}n^\mu = 0$.

The second term of (\ref{22}) is dealt with as follows
\begin{eqnarray}\label{24}
\frac{1}{8} n^\mu e_{i\nu}\nabla_\mu e_{j}^{\ \nu}[\gamma^i , \gamma^j] &=& \frac{1}{8}e_{i\nu} (n^\mu \nabla_\mu e_{j}^{\ \nu} - e_{j}^{\ \mu}\nabla_\mu n^\nu + e_{j}^{\ \mu}\nabla_\mu n^\nu)[\gamma^i , \gamma^j] = \nonumber\\
&=& \frac{1}{8}e_{i\nu} (\mathcal{L}_n e_j)^\nu [\gamma^i , \gamma^j] - \frac{1}{8}e_{i}^{\ \nu} e_{j}^{\ \mu}K_{\mu\nu}[\gamma^i , \gamma^j] \equiv \frac{1}{8}e_{i\nu} (\mathcal{L}_n e_j)^\nu [\gamma^i , \gamma^j] \ .
\end{eqnarray}
To arrive at (\ref{24}), we used the definition of Lie derivative, $(\mathcal{L}_n e_j)^\nu = n^\mu \nabla_\mu e_{j}^{\ \nu}- e_{j}^{\ \mu}\nabla_\mu n^\nu$ as well as the symmetry of the external curvature, $K_{\mu\nu}=K_{\nu\mu}$. Combining (\ref{21}), (\ref{23}) and (\ref{24}) we get
\begin{eqnarray}\label{25}
n^\mu n^\nu \gamma_\mu \nabla^\omega_\nu = \gamma^0 n^\nu \nabla^\omega_\nu =: \gamma^0 \left(\partial_n + \frac{1}{8}e_{i\nu} (\mathcal{L}_n e_j)^\nu [\gamma^i , \gamma^j]\right) + \frac{1}{2}\gamma^i e_{i}^{\ \alpha}\frac{\partial_\alpha N}{N}\ .
\end{eqnarray}
Defining the covariant derivative along the normal, $D_n := -\left(\partial_n + \frac{1}{8}e_{i\nu} (\mathcal{L}_n e_j)^\nu [\gamma^i , \gamma^j]\right)$, we can write the final form of the 3+1 decomposition of the Dirac operator:
\begin{eqnarray}\label{3plus1}
\mathrm{D} = \gamma^0 D_n + {}^{(3)}\mathrm{D} - \frac{1}{2}\gamma^0 K - \frac{1}{2}\gamma^\alpha \frac{\partial_\alpha N}{N}\ .
\end{eqnarray}
The result (\ref{3plus1}) is to be compared with the analogous result from \cite{Sommers:1980mm} where the 3+1 decomposition was achieved using the $SU(2)$ 2-component spinors. The advantage of our calculation is more geometric picture of 3+1 splitting and the possibility to use our result for the Euclidian case where the usage of $SU(2)$-spinors is problematic.


\end{document}